\documentclass[cits,a4paper]{PoS}
\pdfoutput=1
\usepackage{booktabs}
\usepackage{macros}
\usepackage{amsmath}

\title{Non-perturbative quark mass dependence in the heavy-light sector of two-flavour QCD }

\ShortTitle{NP quark mass dependence in the heavy-light sector of two-flavour QCD }

\author{\vskip-1em\includegraphics[scale=0.9]{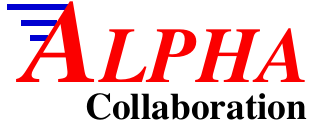}\hfill%
        \begin{minipage}[c]{4cm}\raggedleft\footnotesize%
        \rm\vskip-2em CERN-PH-TH/2008-207\\
            MS-TP-08-25\\
            DESY 08-148\\
            SFB/CPP-08-85
        \end{minipage} \vskip2em
        Michele Della Morte\\
        CERN, Physics Department, TH Unit,
        CH-1211 Geneva 23, Switzerland\\
        E-mail: \email{dellamor@mail.cern.ch}}
        \author{\speaker{Patrick Fritzsch}
        \footnote{Present address: School of Physics and Astronomy, University of Southampton,\newline
        \hspace*{20ex}\! Highfield, Southampton, SO17 1BJ, UK}\;, 
        Jochen Heitger\\
        Westf\"alische Wilhelms-Universit\"at M\"unster, 
        Institut f\"ur Theoretische Physik,\\
        Wilhelm-Klemm-Str. 9, D-48149 M\"unster, Germany\\
        E-mail: \email{fritzsch@uni-muenster.de}, \email{heitger@uni-muenster.de}}
\author{Rainer Sommer\\
        Deutsches Elektronen-Synchrotron DESY, Zeuthen,\\
        Platanenallee 6, D-15738 Zeuthen, Germany\\
        E-mail: \email{rainer.sommer@desy.de}}

\abstract{%
We present preliminary results of the non-perturbative heavy quark mass 
dependence of heavy-light meson observables in the continuum limit of 
finite-volume two-flavour lattice QCD. 
These observables, which are derived from heavy-light Schr\"odinger 
functional correlation functions and computed over a range of 
renormalization group invariant heavy quark masses from the charm to beyond
the bottom region, allow for a quantitative comparison with the predictions 
of HQET and are of practical relevance for solving renormalization problems 
in HQET non-perturbatively by a matching to QCD in finite volume.
}

\FullConference{%
    The XXVI International Symposium on Lattice Field Theory \\
    July 14 - 19, 2008\\
    Williamsburg, Virginia, USA}

\begin{document}
\DeclareFontFamily{T1}{cmr}{\defaulthyphenchar\font=-1}

\section{Introduction}
\label{sec:intro}
The Heavy Quark Effective Theory (HQET) Lagrangian,
  \begin{align}
    \lag{HQET}(x) &= %
    \heavyb(x)\bigg[\,{D_{0}+ m} 
         - {\omega_{\rm kin}}\,\vecD^2 
         - {\omega_{\rm spin}}\,\boldsymbol{\sigma} \vecB 
         \,\bigg]\heavy(x) \,+\,\ord\left( 1/m^2 \right) \,,
    \label{eq:lagrange}
  \end{align}
provides an expansion of QCD amplitudes in the inverse heavy quark mass,
$\minv$, and is renormalizable at any finite order in $\minv$ by
means of power counting. It is a standard phenomenological tool
which simplifies the QCD dynamics in the limit of large masses
like that of the c- or b-quark. 
The HQET Lagrangian as written in \eqref{eq:lagrange} consists of a leading,
static term, $D_0$, which describes the dynamics
in the limit of infinitely heavy 
quark mass, and of the two subleading kinetic and spin terms
whose coefficients are $\omega_{\rm kin}=\omega_{\rm spin}=1/(2m)$
in the classical theory.

To explicitly make HQET an effective theory of QCD requires matching 
calculations to express the parameters in the Lagrangian 
$(m,\omega_{\rm kin},\cdots)$ by those of QCD. 
In principle one could employ perturbation theory in the matching step, 
but owing to the difficulty of a reliable error estimation it may be hard to 
disentangle and quantify deviations coming from higher orders or  
non-perturbative effects in particular, since there are power-divergent 
mixings \cite{Maiani:1992az}.
Despite its phenomenological success --- reflecting, e.g., in determinations 
of $V_{\rm cb}$ \cite{PDG2008,deDivitiis:2008df} or HQET hadronic matrix 
elements where perturbative HQET enters ---, independent and non-perturbative 
tests of HQET may provide a deeper insight into the feasibility of the 
effective theory approach.
In the quenched approximation, such a test was performed and discussed
in~\cite{HQET:pap3}.

In the course of the non-perturbative matching between QCD and HQET, 
originally proposed in \cite{HQET:pap1} and currently being applied to QCD 
with two (massless) dynamical sea quarks~\cite{lat07:hqetNF2} 
(see~\cite{Nf2SF:autocorr} for the variant of the HMC algorithm we use), we here give 
a preliminary report on an extension of the quenched study~\cite{HQET:pap3}
to the two-flavour case.
The non-perturbative large-mass behaviour of some meson observables,
computed in the continuum limit of finite-volume QCD, is confronted with
the static theory in order to investigate the range of validity of the 
$1/m$-expansion and to estimate the size of the $1/m$-corrections.
The use of a \textit{finite} volume of about $(0.5\,{\rm fm})^4$ is 
crucial in this context, since clean non-perturbative comparisons of QCD 
and HQET in the \textit{continuum limit} require $m\ll 1/a$ prior to 
$a\to 0$.

\section{Observables and their large-mass asymptotics}
\label{sec:asympt}
Our observables are built from relativistic heavy-light Schr\"odinger
functional (SF) correlation functions. $\fa$ is a correlator between a
heavy-light pseudoscalar boundary source and an axial current 
operator insertion in the bulk, $\fone$ a boundary-to-boundary correlation 
function and $\kv,\kone$ are their vector channel analogues.
We further use the pseudoscalar, $\fp$, and tensor, $\kt$, correlators 
to improve the bare currents in the respective channels. 
More details and any unexplained notation are found in~\cite{HQET:pap3}.
From now on, these correlators are referred to as their $\ord(a)$ improved 
lattice versions, which amounts to replace the quark bilinears such as the
heavy-light axial current according to  
  \begin{align}
     A_\mu\to \za[1+\tfrac{1}{2}\ba(\amqq{l}+\amqq{h})]\times A_\mu\,.
  \end{align}
We take the non-perturbative values of $\za,\zv$ from 
\cite{impr:za_nf2_pap2,impr:za_nf2} and the 1-loop perturbative estimates 
of~\cite{impr:pap5} for $\bx$, ${\rm X}\in\{ {\rm A},{\rm V} \}$.
Our test observables built from these renormalized correlators are 
  \begin{align}
    \Gamav(L,M) &\equiv   \tfrac{1}{4}\big[\Gamps(L,M)+3\Gamv(L,M)\big]  \,, &
    \Rspin(L,M) &\equiv   \tfrac{1}{4}\ln(\fone/\kone) \,,\\[0.5em] 
    \Yr(L,M)    &\equiv + \dfrac{\fa(T/2)}{\sqrt{\fone}} \,, &
    \Yv(L,M)    &\equiv - \dfrac{\kv(T/2)}{\sqrt{\kone}} \,, \\[0.50em]
    \Raov(L,M)  &\equiv - \dfrac{\fa(T/2)}{\kv(T/2)} \,, &
    \Raop(L,M)  &\equiv - \dfrac{\fa(T/2)}{\fp(T/2)} \,, 
  \end{align}
where the following definitions of the pseudoscalar and vector effective
energies apply:
  \begin{align}
     \Gamps(L,M) &\equiv \left. -\frac{\rmd }{\rmd x_0} \ln\left[\,\fa(x_0)\,\right]\,\right|_{\,x_0=T/2}
                  =             -\frac{\fa'(T/2)}{\fa(T/2)}\,,  \\
     \Gamv(L,M)  &\equiv \left. -\frac{\rmd }{\rmd x_0} \ln\left[\,\kv(x_0)\,\right]\,\right|_{\,x_0=T/2}
                  =             -\frac{\kv'(T/2)}{\kv(T/2)}\,.
  \end{align}
In physically large volume, as $L\to\infty$, $\Gamav$ becomes proportional to 
the spin-averaged mass of the heavy-light meson, $\Rspin$ to the 
spin-splitting term and $\Yr$ ($\Yv$) to the pseudoscalar (vector) heavy-light 
meson decay constant; in this sense we will also use the shorthand decay constant.

These quantities are expected to be described by a 
power series in $1/z$ with logarithmic modifications, where
\begin{align}
        \label{eqn:z-M}
    z &= L_1M \,, &
    M &= \lim_{\mu\to\infty} \left\{ [2b_0\gbsq(\mu)]^{-d_0/(2b_0)}\,\mbar(\mu) \right\}\,,  \\
    \nf&= 2\,,  &
    b_0&= \big(11-\tfrac{2}{3}\nf\big)\big/{(4\pi)^2}\,, \qquad d_0= {8}\big/{(4\pi)^2}  \notag   \,,
\end{align}
and $M$ denotes the renormalization group invariant (RGI) mass of the heavy
quark flavour.
Our choice of lattices with spatial extent $L=L_1$, $T=L$, and the 
corresponding simulation parameters are summarized in 
table~\ref{tab:sim-par}.
In physical units, $L_1$ is about $0.5\,{\rm fm}$. 
For a more detailed account on how the $z$-values under investigation, 
$z\in\{4,6,7,9,11,13,15,18,21\}$, can be kept fixed for given $L_1/a$ in
dependence of the bare parameters, the reader is referred 
to~\cite{lat07:hqetNF2}.

As we are interested in the large $z$-asymptotics of our observables,
one also needs their counterparts computed in the effective theory, as long
as the latter are non-trivial.
We denote the associated quantities by a superscript `stat', 
e.g.~$\fa\to\fastat$. In case of the decay constant we then construct 
\begin{align}
        X(L) &\equiv \frac{\fastat(T/2)}{\sqrt{\fonestat}}   
        \label{}
\end{align}
so that 
\begin{align}
        \lim_{z\to\infty} \Yr(L,M) = X(L) =   \lim_{z\to\infty} \Yv(L,M)
        \label{eqn:Yr-Yv}
\end{align}
holds at the classical level. Due to the heavy-quark spin symmetry in
the static limit, $\Yr$ and $\Yv$ converge to the same limit.

In quantum theory, the scale dependent 
renormalization of the effective theory introduces the mentioned logarithmic 
modifications. 
An example is the axial current in the effective 
theory, where the renormalized 
$X_{\rm R}(L,\mu) = \zastat(\mu) \Xbare(L)$
depends logarithmically on the chosen renormalization scale $\mu$ as well 
as on the renormalization scheme. 
As for the mass, eq.~\eqref{eqn:z-M}, this dependence is removed 
explicitly by passing to the RGI matrix element
\begin{align}
  \XRGI(L) &= \lim_{\mu\to\infty} 
       \big\{ [2b_0\gbsq(\mu)]^{-\gamma_0/(2b_0)}\,X_{\rm R}(L,\mu) \big\}= 
       \zRGIstat\,\Xbare(L)  \,, \qquad \gamma_0= -{1}\big/{(4\pi^2)}\,.
  \label{eqn:XRGI}
\end{align}
For $\nf=2$, the renormalization factor $\zRGIstat$ is known
non-perturbatively from~\cite{zastat:Nf2}.
After expressing our QCD test observables through the corresponding RGIs,
their large-$z$ behaviour is driven by the RGIs of the effective theory 
together with so-called \textit{conversion functions}, $C$, which contain 
the full logarithmic mass dependence of the associated operators.
As arguments of the latter we choose the ratio of RGIs $M/\Lambda$, since
it can be fixed on the lattice without perturbative 
uncertainties \cite{msbar:Nf2}.

The $1/z$-expansions of our test observables now read as follows: 
\begin{align}  \label{eqn:Yx-asymp}
     Y_{\rm X}(L,M) & \;\simas{M\to\infty}\;\;
     \Cx\Big(M/\lMSbar\Big)\,\Big[\XRGI(L)\Big]\,\Big(1 + \rmO(1/z)\Big)\,,  &
     {\rm X}&={\rm PS},{\rm V}, \\\label{eqn:Rpsv-asymp}
     R_{\rm X}(L,M) & \;\simas{M\to\infty}\;\;
     \Cx\Big(M/\lMSbar\Big)\Big[\,1\,\Big]\,\Big(1 + \rmO(1/z) \Big)\;, &
     {\rm X}&={\rm PS/V},{\rm PS/P}, \\\label{eqn:Rspin-asymp}
     R_{\rm spin}(L,M) & \;\simas{M\to\infty}\;\;
     \Cspin\Big(M/\lMSbar\Big)\,\Big[{X^{\rm spin}_{\rm RGI}(L)}\big/z\Big]\,\Big(1 + \rmO(1/z)\Big)\,,
     \\\label{eqn:Gamav-asymp}
     L\Gamav(L,M) & \;\simas{M\to\infty}\;\;
     \Cmass\Big(M/\lMSbar\Big)\Big[\, z \,\Big]\,\Big(1 + \rmO(1/z)\Big)\,.
\end{align}
Beside the overall logarithmic mass dependence governed by the conversion
functions we enclose the leading matrix elements in the effective theory
in square brackets. 
The numerical evaluation of the $\Cx$ proceeds as explained 
in~\cite{HQET:pap3,zastat:Nf2}, where always the highest available 
perturbative approximation of the involved RG functions is employed,
in particular the 3-loop anomalous dimensions $\gamma$ of the axial current
and the chromomagnetic operator in HQET, 
respectively~\cite{ChetGrozin,Grozin:2007fh}. 
By comparing different loop orders for their evaluation, we conclude that 
the remaining perturbative uncertainty of the conversion functions
originating from unknown higher orders is much smaller than the precision of
our lattice data so that a study of the $1/z^n$-corrections becomes feasible.
\begin{table}
 \renewcommand{\arraystretch}{1.25}\small
   \begin{center}
    \begin{tabular}{cccc} \toprule
   $L_1/a$ & $\beta$  &  $\kappa_{\rm l}\approx\hopc$  &  $L_1\ml$  \\\midrule
   $  20$  & $6.1906$ &  $0.135997290$ & $+0.00055(13) $  \\
   $  24$  & $6.3158$ &  $0.135772110$ & $-0.000145(66)$  \\
   $  32$  & $6.5113$ &  $0.135421494$ & $+0.000143(36)$  \\
   $  40$  & $6.6380$ &  $0.135192285$ & $+0.000024(24)$  \\ \bottomrule
    \end{tabular}
   \end{center}
    \vspace{-0.4cm}
  \caption{Simulation parameters that correspond to a line of constant 
  physics characterized by $\gbsq(L_1)=4.484$ and $L_1\ml\approx 0$ in the
  light (i.e.~sea) quark sector. 
  Suitable choices for the hopping parameter of the heavy quark, 
  $\kappa_{\rm h}$, allow to fix its dimensionless RGI mass, $z=L_1M$, to a 
  set of desired values extending from the charm to the bottom quark region.  
  See \cite{lat07:hqetNF2} for details.}
  \label{tab:sim-par}
\end{table}
\section{Results}
\label{sec:results}

  \begin{figure}
    \begin{center}
      \includegraphics[width=0.48\textwidth]{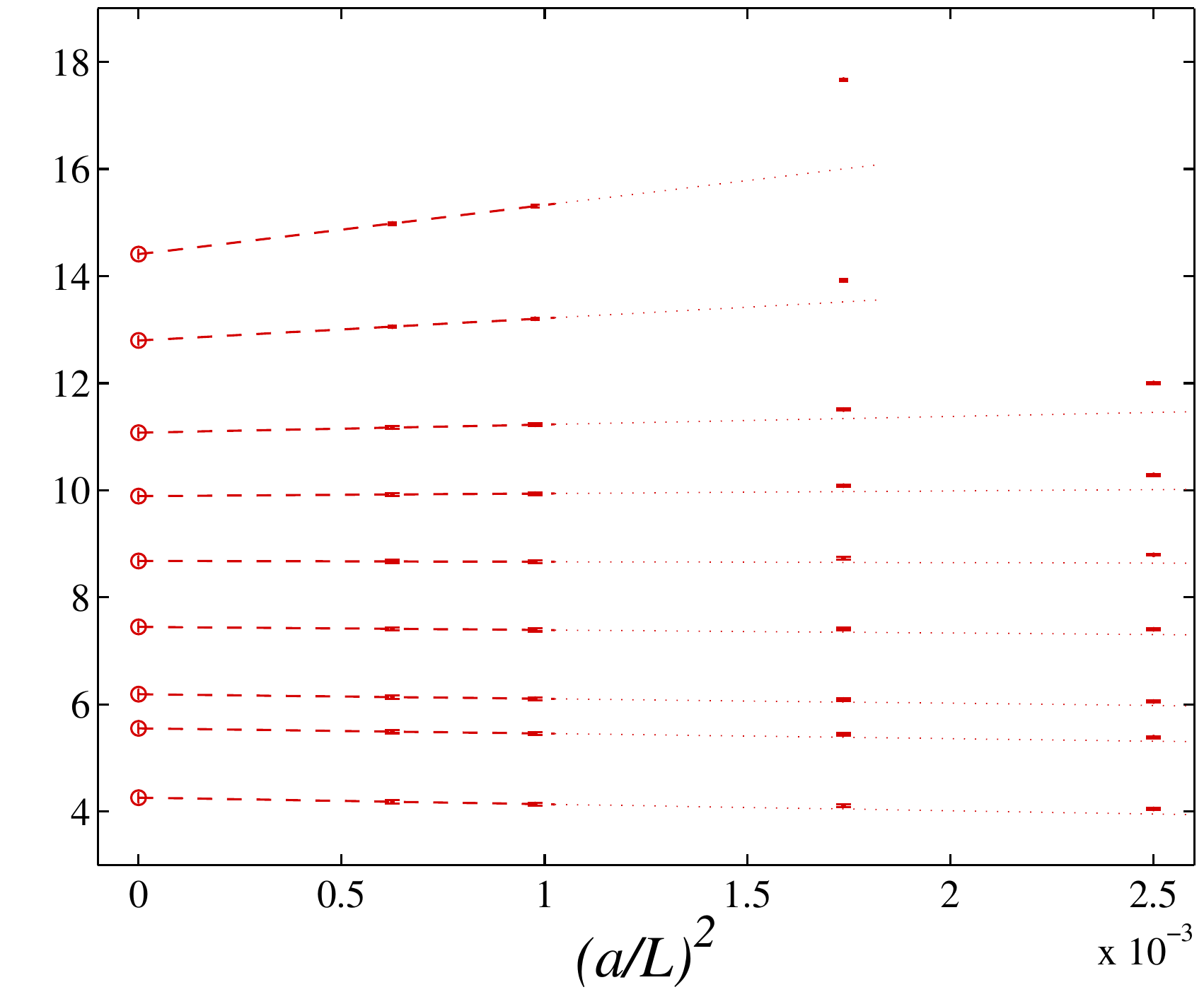}\hfill
      \includegraphics[width=0.48\textwidth]{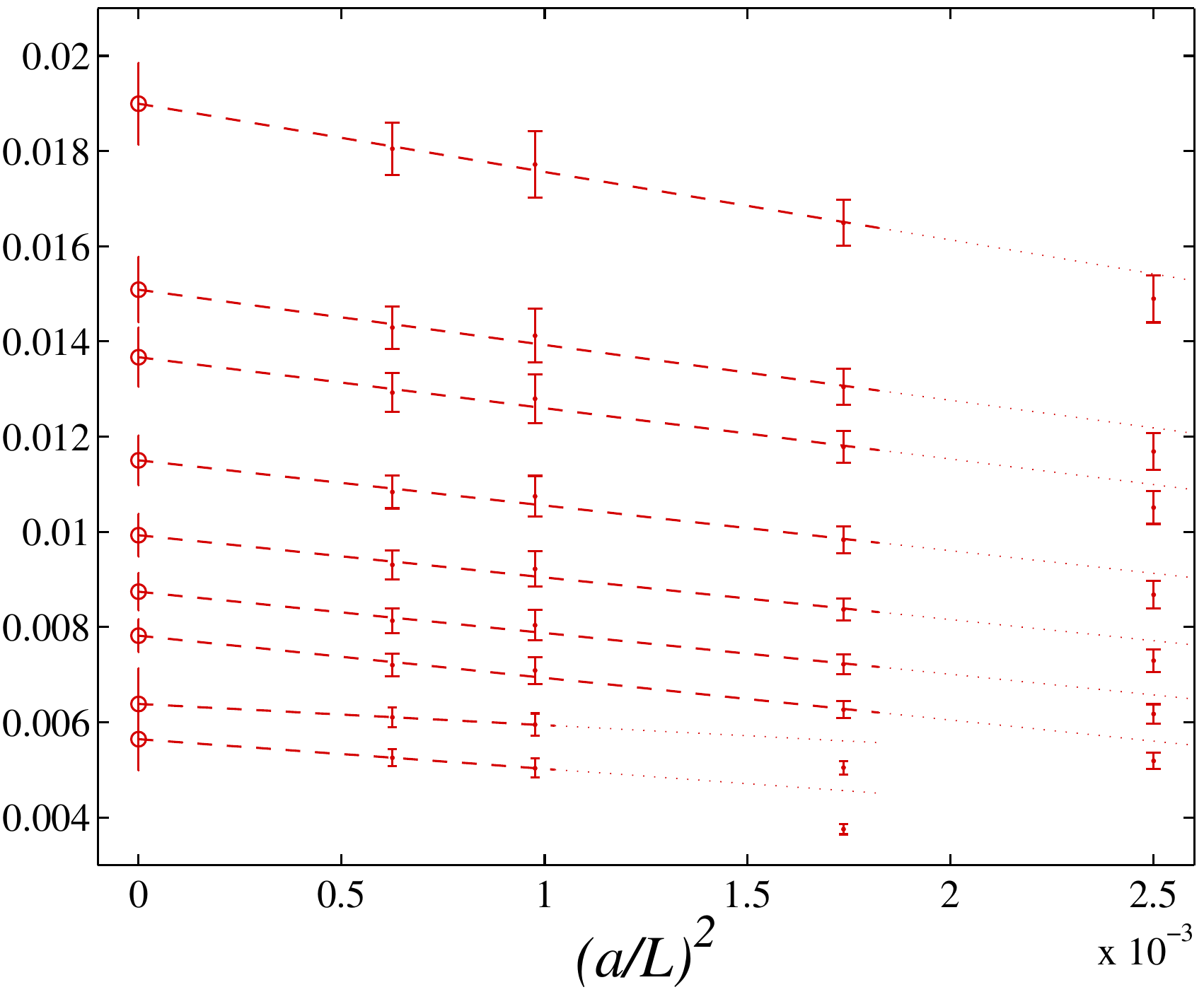}
      \caption{\textit{Left:} Continuum extrapolations linear in $(a/L)^2$
               of the spin-averaged mass $L\Gamav(z,\theta=0.5)$ from $z=4$ 
               (bottom) to $z=21$ (top). 
               The error of the continuum limits is smaller than the symbol 
               size. 
               \textit{Right:} Continuum extrapolations linear in $(a/L)^2$
               of the spin-splitting $\Rspin(z,\theta=0.5)$ from $z=4$ (top) 
               to $z=21$ (bottom).
               --- Depending on the value of $z$, the coarsest
               or the coarsest two lattices are omitted from the fits.}
      \label{fig:LGamma-CL}
    \vspace{-0.5cm}
    \end{center}
  \end{figure}

As examples for the observables under study, continuum extrapolations linear
in $(a/L)^2$ of $L\Gamav$ and $\Rspin$ for the phase parameter\footnote{
The SF boundary conditions in space on the fermion fields are taken to be
periodic up to a global phase $\theta$.
}
$\theta=0.5$ are presented in fig.~\ref{fig:LGamma-CL}. 
In our volume of extent $L=L_1=T\approx 0.5\,{\rm fm}$, which admits to 
reach heavy quark masses up to $M\approx 1.3M_{\rm b}$, the $a\to 0$ 
extrapolations appear to be well controllable provided that one accounts for
the growing (heavy) quark mass in lattice units at given $a/L$ as $z$ is 
increased.
Similar to the quenched work~\cite{HQET:pap3}, this is realized by imposing 
a cut on $aM$ ($aM\lesssim 0.7$) that translates, for given $z$, into the 
coarsest resolutions which may still be included in the continuum 
extrapolations.  
For all observables considered, the growing deviation, as $z$ is increased, 
of the extrapolating fit function from the result at the respective coarsest 
lattice resolution available suggests that between $z=18$ and $z=21$ the
$\ord(a)$ improvement and thus the $a$-expansion has broken down for 
our lattices.

\begin{figure}
    \begin{center}
      \includegraphics[width=0.48\textwidth]{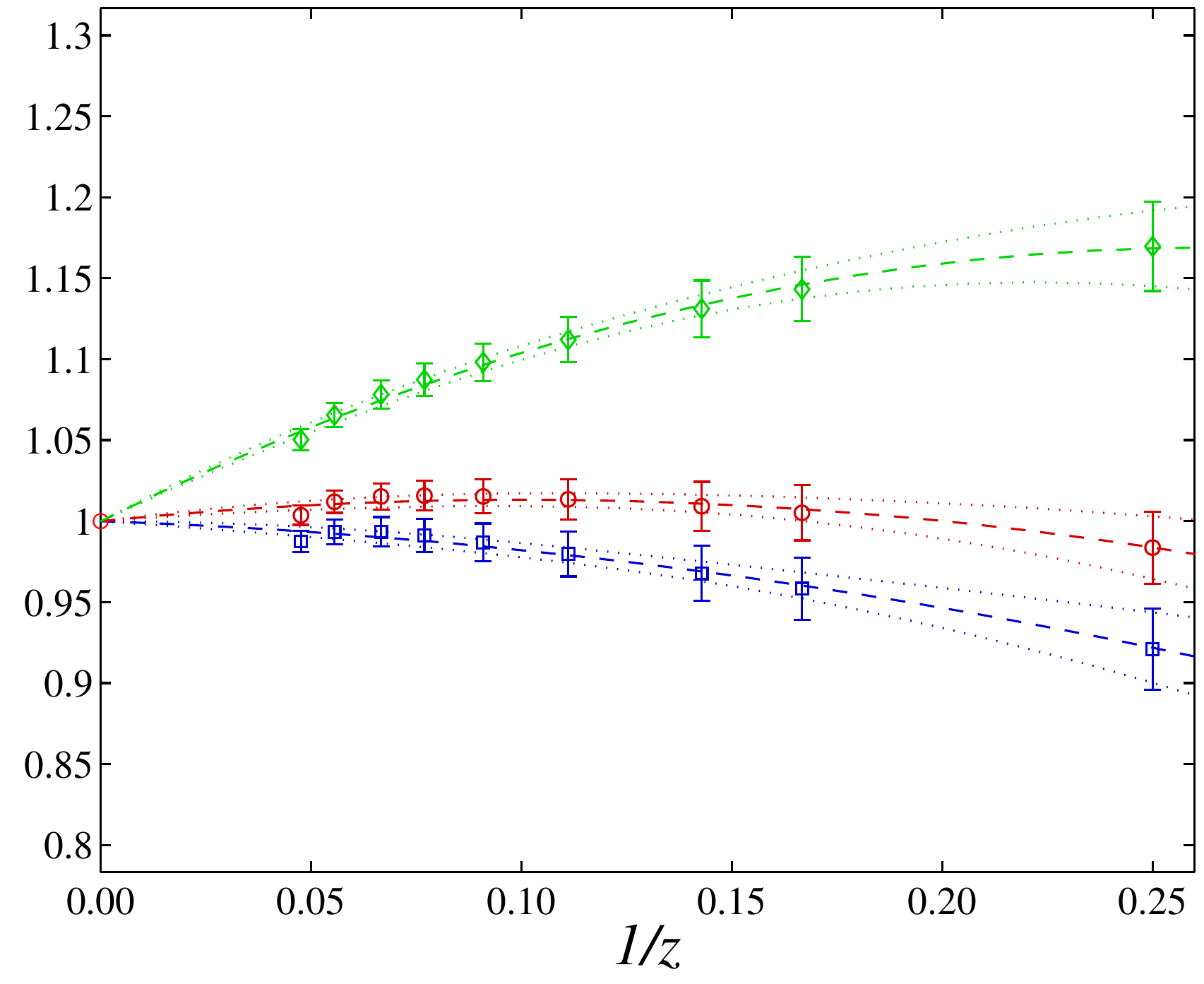}\hfill
      \includegraphics[width=0.48\textwidth]{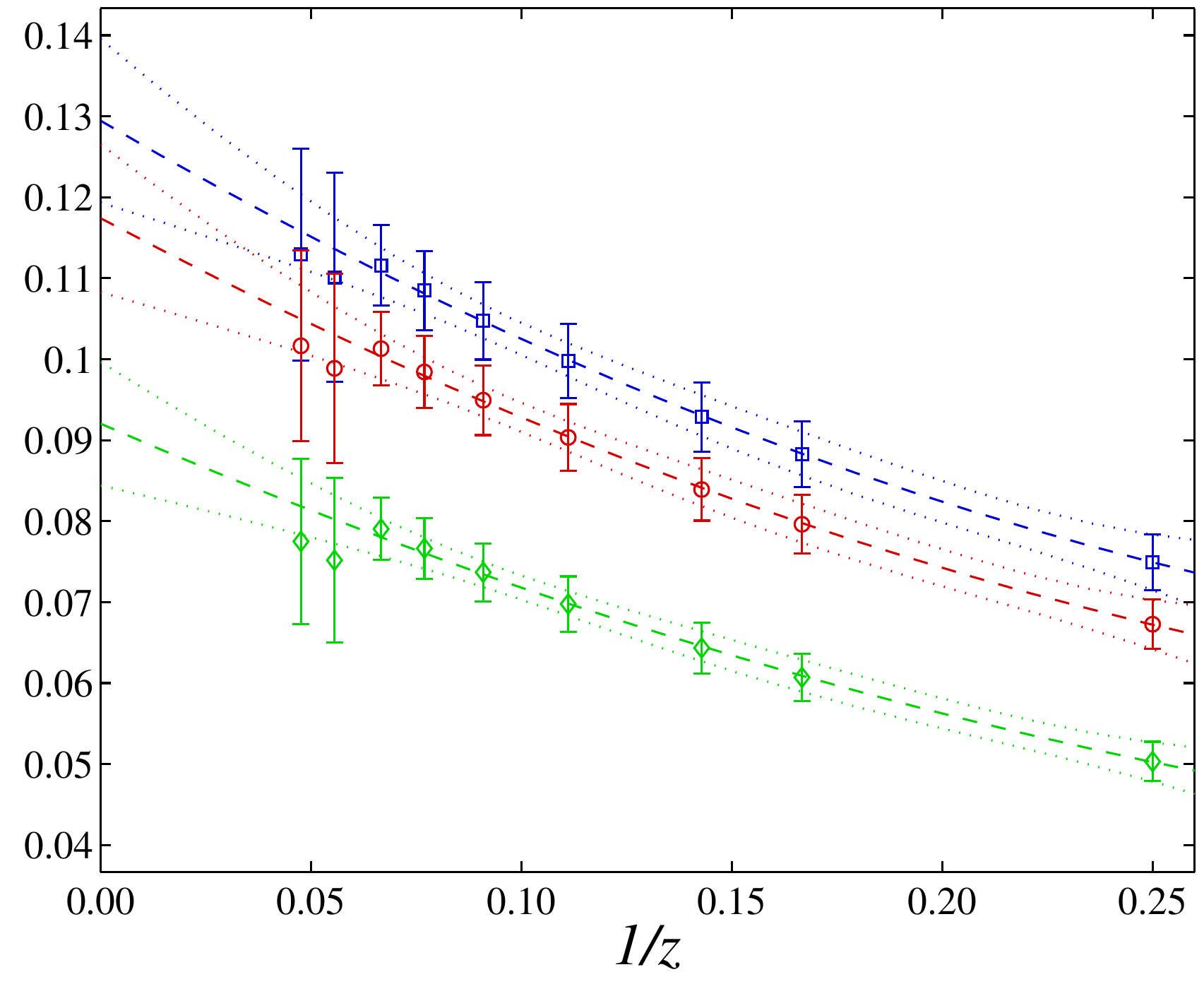}
      \vspace{-0.125cm}
      \caption{\textit{Left:} Asymptotics of $L\Gamav/(z\,\Cmass)$ versus 
               $1/z$.
               All data sets are extrapolated to the (known) static limit
               ($=1$, cf.~\protect\eqref{eqn:Gamav-asymp}) with a constrained 
               quadratic fit. 
               \textit{Right:} $1/z$-dependence of $(z\,\Rspin)/\Cspin$ 
               with an unconstrained quadratic fit to all data points.
               --- Blue, red and green symbols refer to 
               $\theta\in\{0,0.5,1\}$.}
      \label{fig:LGamma-asymp}
    \end{center}
  \end{figure}

\begin{figure}
    \begin{center}
      \includegraphics[width=0.48\textwidth]{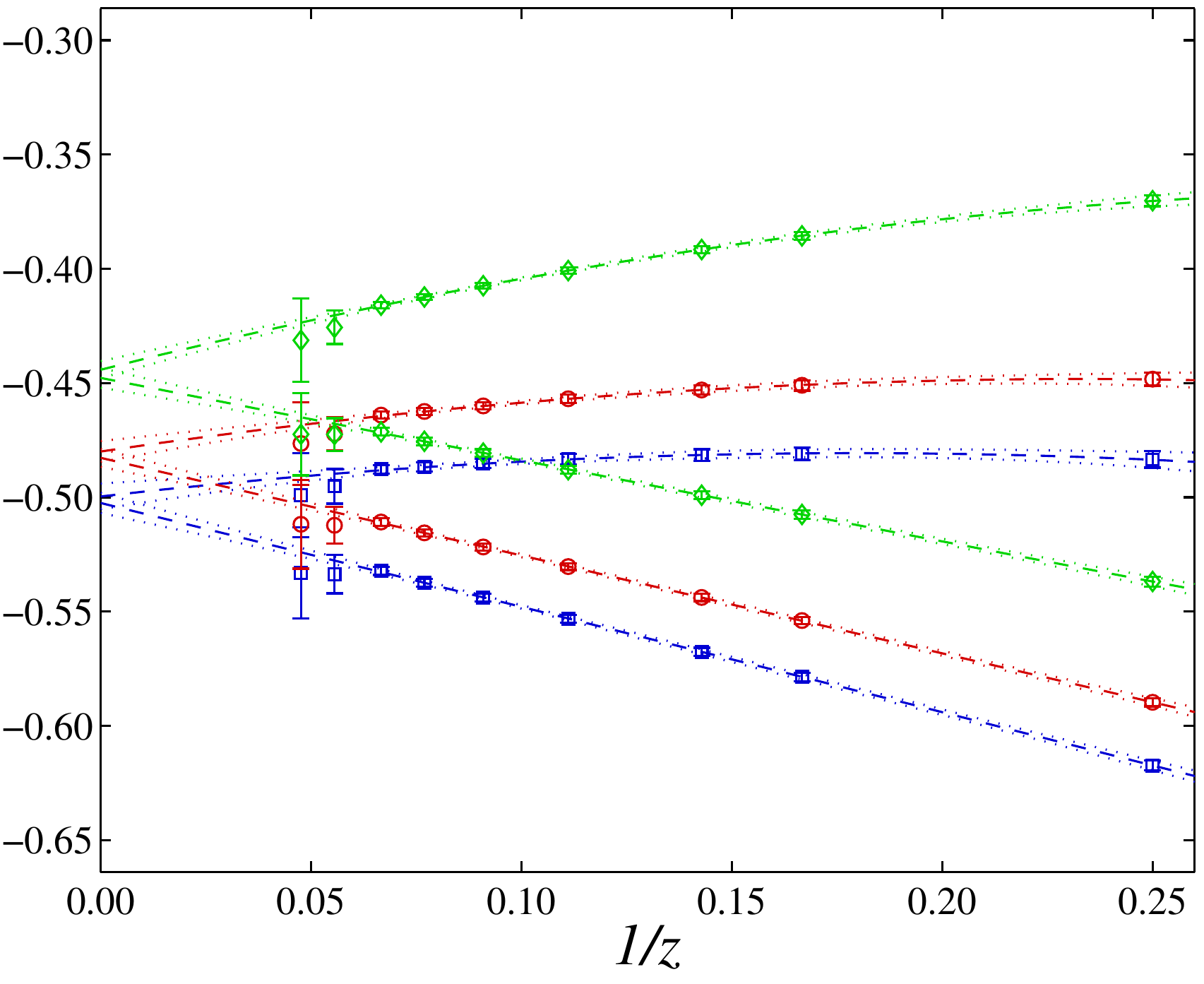}\hfill
      \includegraphics[width=0.48\textwidth]{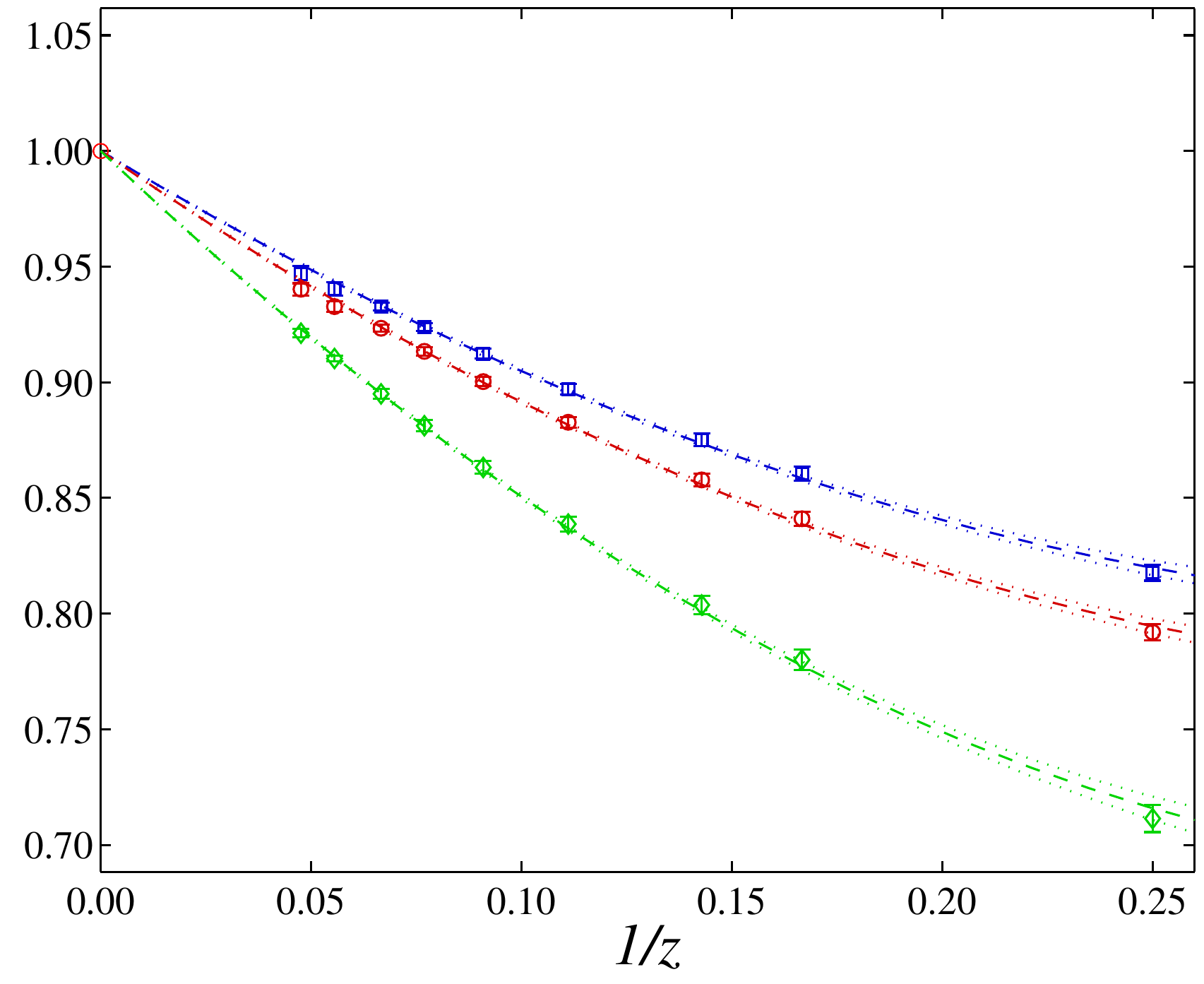}
      \vspace{-0.125cm}
      \caption{\textit{Left:} Asymptotics of $Y_{\rm X}/\Cx$ versus $1/z$
               and its unconstrained quadratic fits to the static limit. 
               Data sets approaching $1/z=0$ from above (below) corresponds 
               to ${\rm X}={\rm PS}$ (${\rm X}={\rm V}$).
               \textit{Right:} $R_{\rm PS/V}/C_{\rm PS/V}$ versus $1/z$ and 
               constrained quadratic fits to the static limit
               ($=1$, cf.~\protect\eqref{eqn:Rpsv-asymp}).}
      \label{fig:decay}
    \vspace{-0.5cm}
    \end{center}
  \end{figure}

Polynomial fits in $1/z$ of the continuum $L\Gamav/(z\,\Cmass)$, 
$(z\,\Rspin)/\Cspin$ as well as of the finite-volume pseudoscalar and vector 
decay constants and their ratio $R_{\rm PS/V}$ (divided by the corresponding
conversion functions) are shown in figs.~\ref{fig:LGamma-asymp} and
\ref{fig:decay}.
If the conversion functions $C_{\rm X}$ are evaluated including the highest 
available perturbative accuracy, the $z$-dependence of $\Gamav$ and 
$R_{\rm PS/V}$ is consistent with the leading term in the $1/z$-expansion, 
which in case of $R_{\rm PS/V}$ is fixed by the heavy-quark spin symmetry of 
the static theory.
Note in addition that the $1/z\to 0$ limit of these observables is 
independent of the choice for the periodicity angle $\theta$
($=0,0.5,1$ here), as it should be.

In general, the $1/z$-corrections of our observables are reasonably small.
As can be inferred from the l.h.s.~of fig.~\ref{fig:decay}, $\Yr/C_{\rm PS}$
and $\Yv/C_{\rm V}$ converge to the same limit (see~\eqref{eqn:Yr-Yv})
as expected.
In principle, this common limit of their $z$-dependence can be constrained 
by the static theory as well, because the renormalization factor of the
associated static axial current matrix element, \eqref{eqn:XRGI}, is already 
known non-perturbatively~\cite{zastat:Nf2}.
The computation of the bare matrix element to obtain the static result for 
$\XRGI$ is in progress.

\section{Conclusions}
We have presented a status report of our ongoing study of the large-mass 
asymptotics of \textit{non-perturbatively renormalized} heavy-light meson 
observables in finite-volume two-flavour QCD.
Their behaviour in dependence of the inverse RGI heavy quark mass is well
compatible with the predictions of HQET confirming the feasibility of a
precise non-perturbative matching of QCD and HQET.
The only perturbative uncertainties owing to the leading logarithmic mass
dependence, which is induced by the conversion functions $C$ relating our 
observables to the RGIs of the effective theory, are under reasonable control.
It appears that for the studied observables the power corrections 
dominate over the perturbative ones 
in the considered range of $z$.

Our continuum extrapolations may be further improved by a removal
of perturbative cutoff effects prior to the extrapolations.
\vskip0.5em
\noindent\textbf{Acknowledgments}\\
This work is part of our effort for precision B-physics with $\nf=2$. 
We would like to thank our colleagues B.~Blossier, G.~de~Divitiis, 
N.~Garron, G.~von~Hippel, R.~Petronzio, H.~Simma and N.~Tantalo for 
discussions and collaboration in this effort.
We thank NIC for allocating computer time on the APE computers to this
project and the APE group at Zeuthen for support. We further acknowledge
partial support by the Deutsche Forschungs\-gemeinschaft under grant 
HE 4517/2-1 and in the SFB/TR 09-03, ``Computational Particle Physics'', 
as well as by the European Community through EU Contract 
No.~MRTN-CT-2006-035482, ``FLAVIAnet''.
P.F. would like to thank the Lattice 2008 organizing committee for 
financial support of his participation in the conference.

\bibliographystyle{JHEP2}
\bibliography{lattice_ALPHA}

\end{document}